\documentclass[
  reprint,
  superscriptaddress,
  aps,
  prl,
  floatfix
]{revtex4-2}

\usepackage{graphicx}
\usepackage{amsmath}
\usepackage{siunitx}
\usepackage{xcolor}

\begin{document}

\title{Self-Similar Multi-Shock Implosions for Ultra-High Compression of Matter}

\author{M. Murakami}
\affiliation{
  Institute of Laser Engineering, Osaka University,
  Suita, Osaka 565-0871, Japan\\
E-mail: murakami.masakatsu.ile@osaka-u.ac.jp}

\date{\today}

\begin{abstract}
We present a class of self-similar solutions describing ultra-high compression
of a uniform-density target by spherically converging, stacked shock waves.
Extending the classical Guderley model, we derive a scaling law for the final
density of the form
$\rho_r/\rho_0 \propto \hat{P}^{\beta(N-1)}$,
where $N$ is the number of shocks, $\hat{P}$ the stage pressure ratio,
and $\beta$ a numerical exponent determined by the adiabatic index $\gamma$.
One-dimensional hydrodynamic simulations confirm the validity of this scaling
across a broad parameter range. Notably, the relation remains accurate
even in the strongly nonlinear regime up to $\hat{P} \sim 70$,
well beyond the perturbative limit, highlighting the robustness and practical
relevance of the model. Owing to its volumetric geometry, this compression
scheme inherently avoids the Rayleigh--Taylor instability, which typically
compromises shell-based implosions, and thereby establishes a theoretical
benchmark for instability-free compression in inertial confinement fusion.
\end{abstract}

\maketitle

\section{I. Introduction}
Self-similar solutions offer analytic insight into nonlinear flows in high-energy-density hydrodynamics.  
A classical example is the Guderley solution~\cite{Gude42, Zeld66}, describing a strong spherically converging shock  
compressing uniform matter (\( \rho_0, P_0 \)) under constant external pressure $P_1$, where $P_1 \gg P_0$.  
It has found broad applications in ICF~\cite{Atze04}, astrophysics, and laser-driven implosions.

The recent ignition demonstration at the National Ignition Facility (NIF)~\cite{Abu_PRL129, Abu_PRL132}  
employed finely tuned multi-shock sequences to optimize core compression,  
highlighting multiple-shock compression as a key strategy in inertial confinement fusion (ICF)~\cite{Robey04, Clark08, Town14}. 
However, current designs remain heavily reliant on numerical optimization,  
with no predictive analytic theory or exact solution available  
for stacked shock convergence in solid targets.  

While the self-similar nature of ablative plasma flows has been explored in low-density regimes—  
notably by Clark \textit{et al.}~\cite{Clark08} in a detailed theoretical study—  
these works primarily focused on thermally driven flows rather than shock-driven compression relevant to inertial confinement fusion.  
In contrast, our study extends the self-similar framework to multi-shock implosions in solid-density matter,  
providing exact analytic solutions and direct validation with hydrodynamic simulations.

In this work, we develop a class of self-similar solutions that generalize the classical Guderley model  
to describe \( N \)-stacked shocks spherically converging toward the center of a solid target.  
The target is subjected to a sequence of external pressures \(  (P_1, P_2, \ldots, P_N) \),  
which are temporally programmed such that all shocks simultaneously arrive at the center.  
The configuration is therefore characterized by a set of stage pressure ratios  $P_i/P_{i-1}$  for $ 2 \le i \le N $.

\section{II. MULTI-STACKED SHOCKS}
We derive analytic scaling laws for the \emph{final reflected density} \( \rho_{\rm r} \),  
defined as the peak density attained immediately after all \( N \) converging shocks merge at the origin  
and reflect outward as a single diverging shock:
\begin{equation}
\frac{\rho_r}{\rho_0} = \tilde\rho_r^\ast  
\left( \prod_{i=2}^{N} \frac{P_i}{P_{i-1}} \right)^\beta = 
\tilde\rho_r^\ast \left(\frac{P_N}{P_1}\right)^\beta.
\label{eq:general_rho}
\end{equation} 
Here, \( \tilde\rho_{\rm r}^\ast = \rho_{\rm r}^\ast/\rho_0 \) is the reflected density compression ratio  
for a single converging shock, with the asterisk “\( \ast \)” denoting the baseline Guderley solution (\( N = 1 \)).  
The numerical exponent \( \beta(\gamma) \), analytically derived via a first-order perturbation  
around this solution, is formally valid in the weakly nonlinear limit.

For planar geometry, it can be analytically shown using a chain of Rankine--Hugoniot relations  
that the maximum density under \( N \)-stacked shock waves is achieved  
when the applied pressures form a geometric progression, i.e.,  
\[
\frac{P_2}{P_1} = \frac{P_3}{P_2} = \cdots = 
\left(\frac{P_N}{P_1}\right)^{1/(N-1)} \equiv \hat{P},
\]
where \( \hat{P} \) is the specific stage pressure ratio, which is kept constant over all the pressure stages, $1\le i\le N$.  
In the following spherical analysis, we initially adopt this geometric configuration for simplicity.  
Remarkably, we later find that the resulting scaling law is not only applicable,  
but also unexpectedly robust and broadly valid.

In the special case where the successive pressures form a geometric sequence
with a constant stage ratio $\hat{P} \equiv P_i/P_{i-1}$,
Eq.~(1) reduces to a simple power-law scaling,
$\rho_r / \rho_0 \propto \hat{P}^{\beta(N-1)}$,
and the cumulative ratio is $P_N/P_1 = \hat{P}^{\,N-1}$.
However, the compression saturates in the strong-shock regime  
as it approaches the asymptotic limit set by the single-shock solution.
Accordingly, the final density compression ratio follows the piecewise form:
\begin{equation}
\frac{\rho_r}{\rho_0} =
\begin{cases}
\tilde\rho_r^\ast\, \hat{P}^{\beta(N-1)} & \text{for } \hat{P} \lesssim (\tilde\rho_\infty^\ast)^{1/\beta}, \\
\tilde\rho_r^\ast\, (\tilde\rho_\infty^\ast)^{N-1} & \text{for } \hat{P} \gtrsim (\tilde\rho_\infty^\ast)^{1/\beta},
\end{cases}
\label{eq:rho_r_scaling}
\end{equation}
where \( \tilde\rho_\infty^\ast = \rho_{\infty}^\ast / \rho_0 \)  
is the asymptotic compression ratio in the strong-shock limit.  
The critical pressure ratio  
\( \hat{P}_c \equiv (\tilde\rho_\infty^\ast)^{1/\beta} \)  
is obtained by equating the two branches and marks the boundary  
between the perturbative and asymptotic regimes.
Note that, for planar geometry, the asymptotic limit is given by  
\( \tilde\rho_\infty = (\gamma + 1)/(\gamma - 1) \)~\cite{Zeld66}.

Thus, for the representative case of an ideal gas with \( \gamma = 5/3 \),  
we obtain \( \tilde{\rho}_{\rm r}^\ast = 32.3 \), \( \tilde\rho_\infty^\ast = 9.55 \)~\cite{Gude42, Zeld66}
as  the baseline parameters (\( N = 1 \)),  
and \( \beta = 0.724 \).  
Under these values, Eq.~(2) reduces to
\begin{equation}
\frac{\rho_r}{\rho_0} \simeq
\begin{cases}
32.3 \cdot \hat{P}^{0.724(N-1)} & \text{for } \hat{P} \lesssim 23, \\
32.3 \cdot 9.55^{N-1} & \text{for } \hat{P} \gtrsim 23.
\end{cases}
\label{eq:rho_r_gamma53}
\end{equation}

{\color{black}
The areal mass density (or areal density $\rho R$) of the reflected shock,  
$\langle\rho r\rangle_r \equiv \int^{R(t)}_0 \rho(t,r)\,dr,$
is a key parameter in determining the fuel burn fraction in ICF. The self-similar solution gives the reflected-shock density profile in the approximate form  $\rho(r,t)/\rho_r \simeq \left(r/R(t)\right)^{0.7}$ for $r \le R(t)$.
Assuming the expanding  front at \(r=R(t)\) originates from a fluid element initially located at \(r=r_0\), the mass conservation condition,  
$\frac{4\pi}{3}\rho_0 r_0^3 = \int^{R(t)}_0 4\pi r^2 \rho(r,t)\,dr$, 
yields the approximate relation  
\[
\frac{\langle\rho r\rangle_r}{\rho_0 r_0} \simeq 0.63\left(\frac{\rho_r}{\rho_0}\right)^{2/3},
\tag{4}
\]  
where the numerical prefactor 0.63 arises from the assumed similarity profile.
For instance, taking \(\langle\rho r\rangle_r=1-2\,\,\text{g/cm}^2\) as the requirement for achieving medium burn fraction,  \(r_0 \approx 400 - 800~\mu\text{m}\) assuming $\rho_r/\rho_0=2000$ and \(\rho_0 = 0.25\ \text{g/cm}^3\) for liquid DT. This scaling clearly demonstrates that ignition-relevant  
areal densities can be attained with practical target dimensions,  
underscoring the feasibility of multi-shock-driven designs. 
This requirement also maps directly onto the corresponding pair 
of $\hat{P}$ and $N$, as determined from Eq.~(3).
}

\begin{figure}
\centering
\includegraphics[width=75mm]{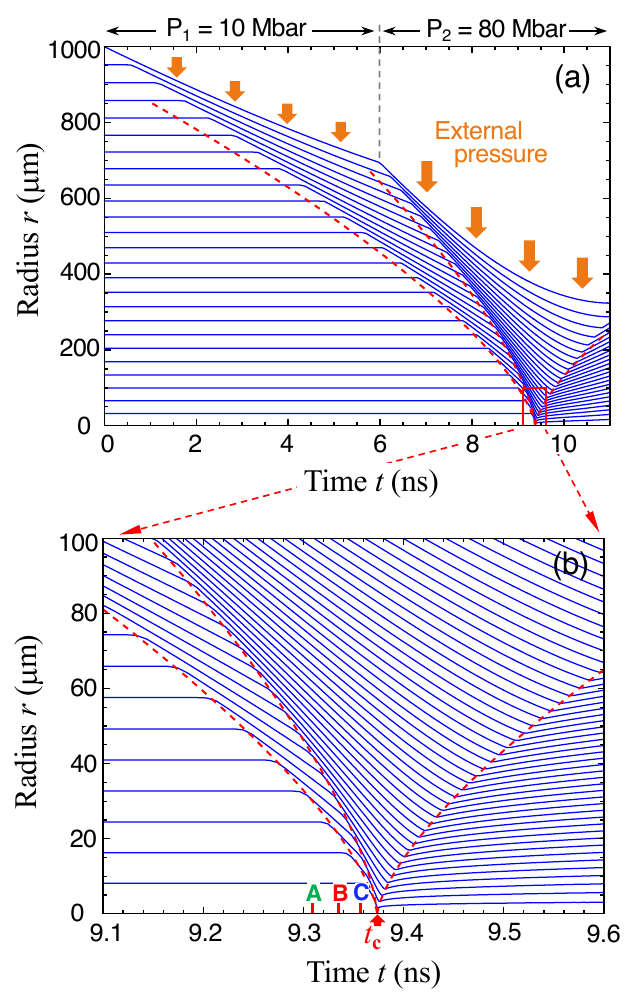}
\caption{
(a) Lagrangian trajectories (blue) showing the 
implosion of a solid DT sphere  
under two-stage pressure loading with $\hat{P} = 8$.
{\color{black}
Dashed lines: self-similar theoretical trajectories (from panel (b)) over-plotted for direct comparison with the simulations.
}
Shocks are synchronized to collide simultaneously at the center at $t=t_c$.  
(b) Close-up view near $t = t_c$, where both shocks obey the scaling $r \propto |t - t_c|^\alpha$.
}
\label{fig1}
\end{figure}

\section{III. HYDRODYNAMIC SIMULATION}
We tested our analytic model by conducting one-dimensional Lagrangian hydrodynamic simulations 
of multi-shock implosions in spherical geometry. 
Figure~1(a) shows the fluid trajectories (blue) resulting from a two-stage pressure sequence, 
with $P_1 = 10$~Mbar followed by $P_2 = 80$~Mbar, corresponding to a stage pressure ratio 
of $\hat{P} = 8$.
The shocks are precisely timed to converge simultaneously at the center. 
{\color{black}
For visual guidance, Fig.~1(a) superimposes the analytic self-similar trajectories
(red, extrapolated from the similarity solution) on the simulated shock loci across the entire
converging stage, showing close agreement up to the collapse.
A zoomed view in Fig.~1(b) highlights the final collapse, where the red trajectories follow
the expected self-similar scaling \( r \propto |t - t_c|^{\alpha} \), confirming the underlying
similarity structure.
}

\begin{figure}
\centering
\includegraphics[width=80mm]{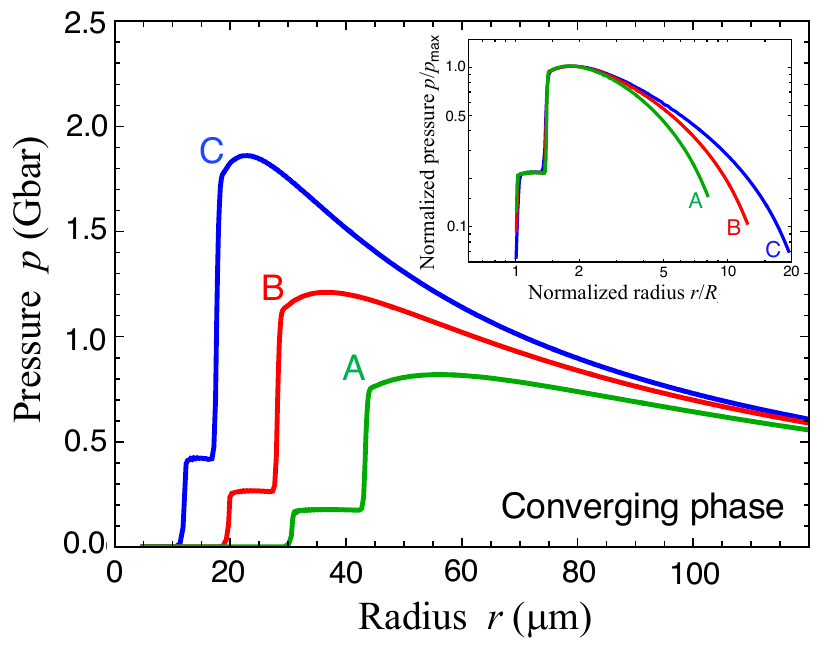}
\caption{Pressure profiles at three times (A, B, C) during convergence,  
corresponding to markers in Fig.~1(b).  
The peak pressure rises and the shock sharpens, preserving the two-step structure.  
Inset: Log–log scaling collapse confirms self-similar evolution.
{\color{black}
Minor deviations at large radii reflect the pre-asymptotic flow, which shrinks as the collapse proceeds.
}
 }
\label{fig2}
\end{figure}

Figure~2 presents pressure snapshots taken at three representative times labeled A, B, and C in Fig.~1(b). When normalized by the instantaneous first shock radius and peak pressure, the profiles collapse onto a single curve in the log--log inset, providing direct evidence of the self-similar evolution. The preservation of the two-shock structure during the entire convergence process further supports the staged compression mechanism.
It should be noted that in a spherically converging shock, self-similarity is an asymptotic property in time: the similarity is first established in the vicinity of the accelerating shock front, while the outer flow approaches the similarity attractor only gradually as $t\rightarrow t_c$. Thus, the minor deviations visible at large radii in Fig. 2 reflect the pre-asymptotic outer region. As the collapse proceeds, the similarity domain expands and ultimately fills the interior region (see, e.g., $\S13.3$ of Ref.[2]).

\section{IV. SELF-SIMILAR SOLUTION}
To quantify this mechanism, we begin by recalling the classical Guderley solution~[1],  
which describes a self-similar implosion of a single converging shock (\( N = 1 \)) in an ideal gas. 
This solution is characterized by a similarity exponent \( \alpha^\ast \),  
determined as an eigenvalue of the reduced hydrodynamic equations.  
For \( \gamma = 5/3 \), the exponent takes the value \( \alpha^\ast = 0.68838 \). 
We then extend this framework to form stacked converging shocks.

The converging flow of such stacked shocks can be examined by analyzing  
solution trajectories in the $(U, Z)$ phase space~\cite{Zeld66},  
where $U$ and $Z$ denote the dimensionless fluid velocity and sound speed squared, respectively.  
These trajectories arise from reducing the original PDE system via a similarity ansatz,  
leading to a compact set of ODEs that are solved as an eigenvalue problem.

Figure~3(a) shows the integrated solution trajectories for the two-shock case ($N = 2$).  
The blue curve starts at point 1, corresponding to the first shock front,  
and proceeds smoothly to point 2.  
Point 2 is not uniquely specified, as the position of the second shock front  
can be freely chosen within a physically admissible range.  
As a result, the self-similar solution admits a continuous 
family of flow profiles,  
each corresponding to a different choice of point 2.  
{\color{black}
However, once the global requirement is imposed that all shocks must converge simultaneously at the origin, this apparent freedom is effectively removed and the physically relevant trajectory is uniquely determined.
}
In Fig.~3(a), the coordinate of point 2 is fixed by comparison with  
the hydrodynamic simulation results shown in Figs.~1 and 2.  
Between points 1 and 2, the solution can cross the so-called sonic line,  $Z = (U - \alpha^\ast)^2$, smoothly only when the eigenvalue takes the specific value $\alpha^\ast = 0.68838$.

\begin{figure}
\centering
\includegraphics[width=86mm]{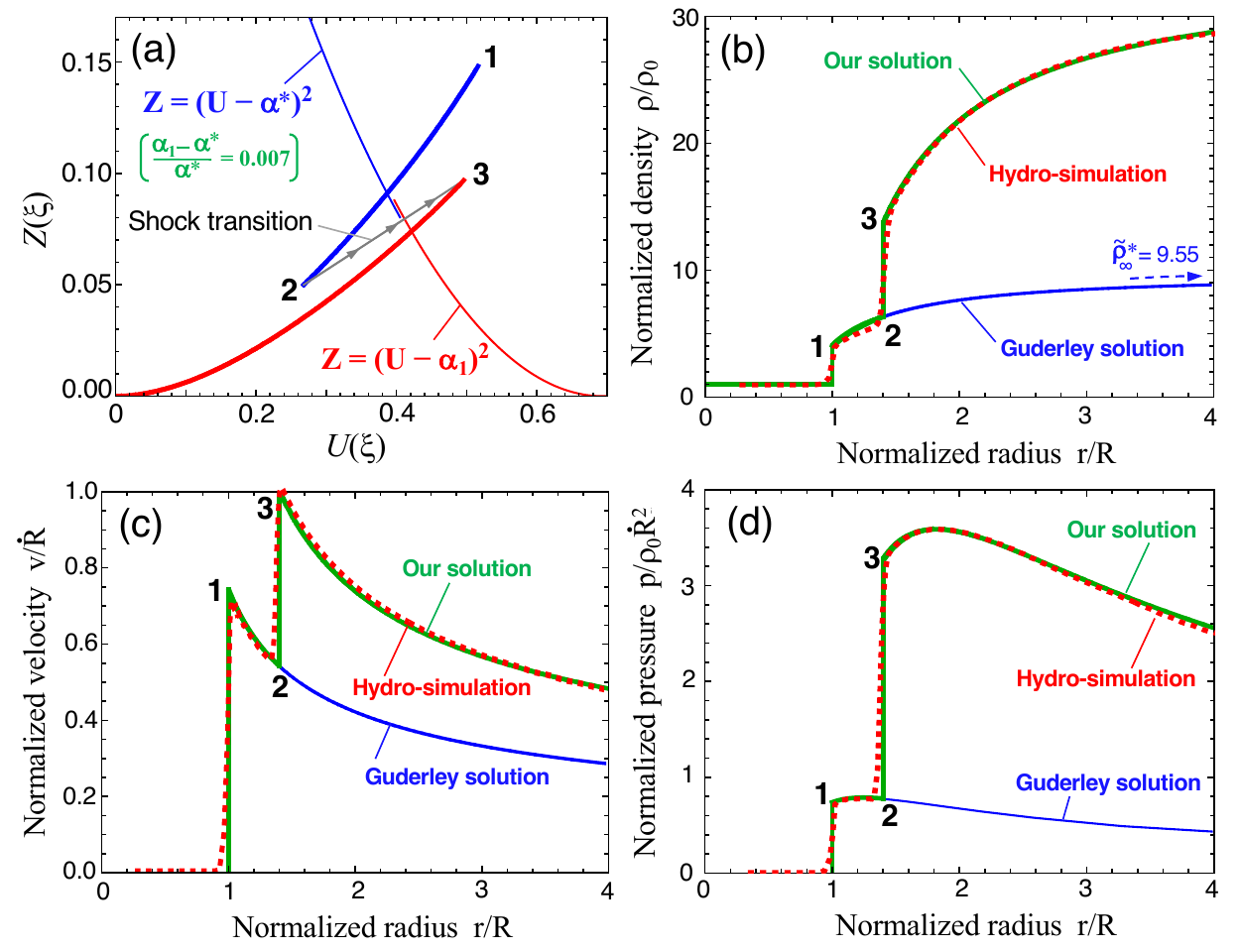}
\caption{
(a) Integrated trajectories in the U--Z plane for $\gamma = 5/3$, 
showing the first-shock (blue, $\alpha^\ast = 0.68838$) and 
second-shock (red, $\alpha_1 = 0.69345$) solutions. 
{\color{black}Thin solid curves indicate the sonic lines $Z = (U - \alpha)^2$.} 
(b)--(d) Scaled radial profiles of density, velocity, and pressure during convergence. 
Solid lines: hydrodynamic simulations; dashed lines: self-similar solutions, 
showing excellent agreement.
}
\label{fig3}
\end{figure}

At point 2, the trajectory undergoes a discontinuous compression to point 3,  
corresponding to the second shock front, forming a jump that satisfies the Rankine–Hugoniot condition.  
For the second shock trajectory shown in red, the solution passes smoothly through the sonic line;  
however, a singularity still arises at the sonic point.  
To remove this singularity and ensure a regular solution at the origin,  
a slightly different eigenvalue, \( \alpha_1 = 0.69345 \), is required.
{\color{black}
This value arises from the same eigenvalue condition as in the Guderley case, namely, the requirement that the similarity trajectory passes smoothly through the sonic line without singularity. Because the trajectory after the Rankine–Hugoniot jump (point 2 → point 3) imposes a modified boundary condition, the resulting eigenvalue differs slightly from the original one, yielding $\alpha_1=0.69345$.
}

This correction in the eigenvalue leads to a modified self-similar flow structure,  
as shown in Figs.~3(b–d), where the numerical hydrodynamic simulations  
are compared with the analytic similarity solutions for both \( N = 1 \) and \( N = 2 \).  
The excellent agreement is observed not only in the overall profiles but also in key physical fields—  
density, velocity, and pressure—capturing both the near-origin density amplification and  
the structure of the second shock.  
These results validate our analytic extension of the Guderley framework  
to multi-shock configurations and underpin the physical basis of the scaling law introduced in Eq.~(3).

\begin{figure}
\centering
\includegraphics[width=86mm]{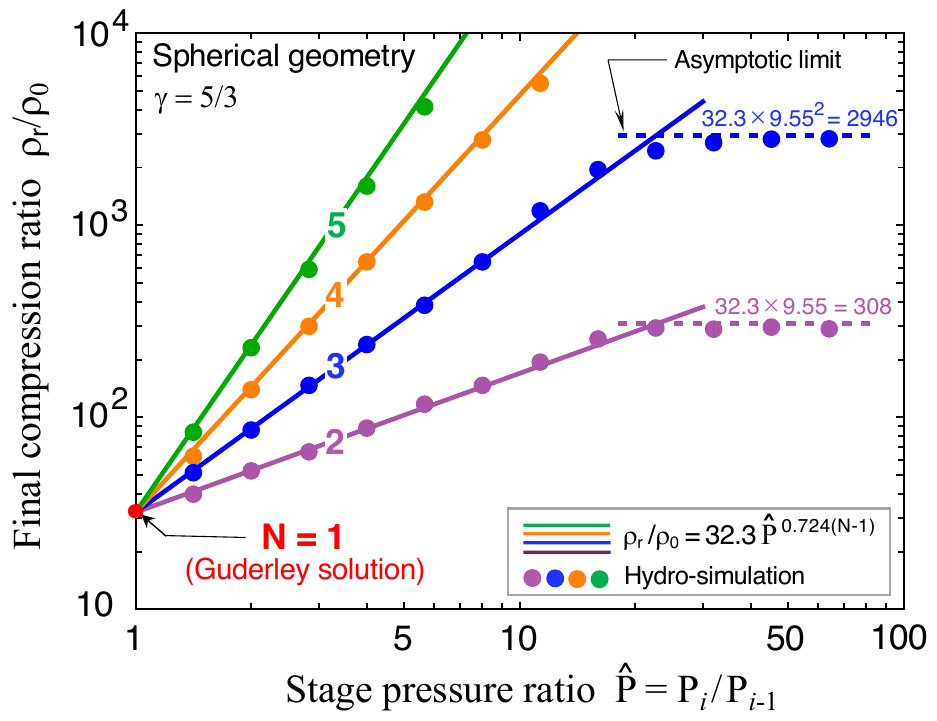}
\caption{ Final compression $\rho_r/\rho_0$ versus stage pressure ratio $\hat P$ for $N=1$–5 stacked shocks. 
{\color{black}Curves 1–5 denote $N=1$–5 (1 = single-shock/Guderley).
}
$\rho_r$ is the reflected-shock crest density immediately after on-axis coalescence and is subsequently advected outward with the diverging shock. Dashed lines indicate the asymptotic limits.
}
\label{fig4}
\end{figure}

\section{V. Compression Scaling}
Figure~4 shows the final compression ratio \( \rho_r/\rho_0 \)  
as a function of the stage pressure 
ratio \( \hat{P} \) for various \( N \).  
{\color{black}
Here, $\rho_r$  is defined as the peak density at the crest of the reflected shock immediately after coalescence at the center; thereafter the crest is advected with the expanding shock and remains essentially conserved—a hallmark of the diverging phase.
}
The simulation data (colored dots) exhibit excellent agreement with the analytic prediction, Eq.~(3),  
across a remarkably wide range (\( \hat{P} \sim 1 \)--70), encompassing both the perturbative (\( \hat{P} \lesssim 23 \))  
and strongly nonlinear (\( \hat{P} \gtrsim 23 \)) regimes.  
In the weak-shock regime, the perturbative scaling holds with high fidelity,  
while in the strong-shock regime, the results asymptotically approach the theoretical limit  
set by the Guderley similarity solution and the Rankine--Hugoniot relation.

This broad agreement suggests that geometric focusing 
in spherical implosions  
not only enhances compression but also acts as a self-regulating mechanism  
that suppresses nonlinear distortions during convergence.
{\color{black}
This interpretation is indirectly supported
by Fig.~4, where the simulation data remain in close agreement
with the analytic scaling law (Eq.~3) across both perturbative
and strongly nonlinear regimes. Significant nonlinear distortions
would otherwise be expected to manifest as systematic deviations
from the theoretical curve, which are not observed.
}

A key structural advantage of this approach lies in its volumetric nature:  
energy is delivered in discrete stages and focused geometrically 
toward the center,  
inherently suppressing the Rayleigh--Taylor (R--T) instability that typically compromises compression  
in shell-based implosions.  
This intrinsic immunity to hydrodynamic instability, combined with full analytic tractability,  
establishes a promising pathway to high-efficiency, instability-free compression in high-energy-density systems.  
The derived scaling, \( \rho_r/\rho_0 = 32.3 \times 9.55^{N-1} \), remains valid even in the strongly nonlinear regime  
\( \hat{P} \gtrsim 23 \), highlighting the robustness and wide applicability of the model.

\section{VI. Discussion}
{\color{black}
It should be noted that, unlike thin-shell implosions, shocking a uniform solid to ignition-relevant densities requires substantially higher drive pressures. 
This drawback, however, can be alleviated by the present multi-shock self-similar scheme, in which the stage pressure requirement is systematically reduced with increasing shock number $N$ (see Fig.~4). 
While we do not claim immediate competitiveness with the optimized thin-shell implosions realized at the National Ignition Facility, the present framework provides a complementary paradigm: a volumetric, Rayleigh–Taylor–immune implosion with fully analytic tractability, which highlights a principled route to instability-free compression.
We also note that increasing $N$ effectively drives the compression toward a quasi-isentropic limit, reminiscent of classical ramp-compression models \cite{Kidd76}, while retaining the practical robustness of shock-driven, self-similar focusing.
}

From a practical standpoint, implementing such stacked-shock schemes  
requires precise synchronization of drive pulses.  
While achieving this level of control remains experimentally challenging,  
recent progress in pulse shaping and timing diagnostics continues to enhance the feasibility of such designs.  
The recent achievement of ignition at the National Ignition Facility (NIF)  
underscores the importance of precise shock orchestration in achieving high compression~\cite{Abu_PRL129}.  
Our results provide a complementary theoretical framework that elucidates the scaling behavior  
of multi-shock-driven implosions and offers analytic guidance for optimizing future target designs.

The critical importance of shock timing is also emphasized in the shock ignition approach to inertial confinement fusion~\cite{Bett16}, 
where a final high-intensity shock is launched near stagnation to ignite the compressed core. 
In contrast, our framework relies on multiple, geometrically programmed shocks that synchronize at the collapse point, 
enhancing compression rather than ignition. 
Both approaches underscore the need for precise temporal control of shock propagation.

The present framework may also find application in alternative fusion concepts  
where ultrahigh compression is essential, such as proton--boron fusion~\cite{Marg22}.  
Achieving densities in the range of \(10^3\)--\(10^4\) times the solid density  
is often considered favorable in such scenarios,  
and the stacked-shock approach offers a physically grounded and potentially realizable pathway  
to reach these regimes.

While our analysis focuses on the density scaling achieved through temporally stacked shocks, 
we note that prior theoretical and numerical studies have demonstrated a scaling of the stagnation pressure in spherical implosions 
as \( p_s / p_0 \propto M_0^3 \) under ideal conditions~\cite{Kemp01}. 
Although this pressure scaling is not explicitly derived here, it reflects the same underlying mechanism—namely, 
the collective effect of converging shocks—that governs our density scaling model. 
Our results thus offer a complementary analytic approach that may help unify density- and pressure-based scaling perspectives 
in the design of high-performance implosions.

\section{VII. SUMMARY}
In summary, in this study, we have developed a class of self-similar solutions that  
extend the classical Guderley model to describe implosions driven by  
multiple converging shocks. The resulting scaling laws provide a  
unified analytic description of compression performance, bridging  
both perturbative and asymptotic regimes. Their validity is  
confirmed by hydrodynamic simulations across a broad range of  
pressure ratios. These findings establish a robust theoretical  
foundation for volumetric, instability-free implosion designs, and  
suggest promising directions for high-efficiency compression schemes  
in inertial confinement and advanced fusion scenarios, including  
proton--boron fusion~\cite{Marg22}.

\section*{Acknowledgments}

Supported by JSPS.  
The author thanks Prof. J.~Meyer-ter-Vehn for insightful discussions that guided the theory.

\end{document}